\documentclass[aps,prd,groupedaddress]{revtex4-2}
\usepackage{graphicx}
\usepackage{epsfig}
\usepackage{amsfonts}
\usepackage{amssymb}
\usepackage{epsf}
\usepackage{color}
\usepackage[utf8]{inputenc}
\newcommand{\insertplot}[5]{\begin{figure}
 \hfill\hbox to 0.05in{\vbox to #5in{\vfill
 \inputplot{#1}{#4}{#5}}\hfill}
 \hfill\vspace{-.1in}
 \caption{#2}\label{#3}
 \end{figure}}
\newcommand{\inputplot}[3]{
 \special{ps: plotfile #1}
}

\newcounter{fig}

\begin{document}

\title{Scalarized Black Holes}
\author{Jose Luis Bl\'azquez-Salcedo}
\email[]{jlblaz01@ucm.es}
\affiliation{Departamento de F\'isica Te\'orica II and IPARCOS, Facultad de Ciencias F\'isicas, 
Universidad Complutense de Madrid, 28040 Madrid, Spain}
\author{Burkhard Kleihaus}
\email[]{b.kleihaus@uni-oldenburg.de}
\affiliation{Institute of Physics, University of Oldenburg, D-26111 Oldenburg, Germany}
\author{Jutta Kunz}
\email[]{jutta.kunz@uni-oldenburg.de}
\affiliation{Institute of Physics, University of Oldenburg, D-26111 Oldenburg, Germany}

\date{\today}% It is always \today, today,
\begin{abstract}
Black holes represent outstanding astrophysical laboratories to test the 
strong gravity regime, since alternative theories of gravity may predict 
black hole solutions whose may differ distinctly from those of General Relativity. 
When higher curvature terms are included in the gravitational action as, 
for instance, in the form of the Gauss-Bonnet term coupled to a scalar field, 
scalarized black holes result.
Here we discuss several types of scalarized black holes
and some of their properties.
\end{abstract}

\maketitle

\section{Introduction}

The existence of black holes in the Universe, following gravitational collapse,
is a genuine prediction of General Relativity (GR) \cite{Penrose:1964wq}.
However, in GR their properties are highly constrained,
when one assumes the standard model of particle physics
for the allowed matter fields
and considers astrophysically relevant black holes.
The expectation of 
astrophysical black holes being (basically) uncharged,
then leads to the conclusion that 
they are (in good approximation) all described by the Kerr family of rotating black holes,
i.e.,  asymptotically flat black hole solutions of the vacuum Einstein equations.

Kerr black holes are uniquely characterized by their mass and their
angular momentum (see, e.g. \cite{Chrusciel:2012jk})
and thus they carry \textit{no hair}.
All their multipole moments are given in terms of these two quantities.
Also, Kerr black holes are subject to a bound on their angular momentum,
%$j = \frac{J}{M^2} \le  1$, 
which is reached in the extremal limit.
Beyond this bound only naked singularities reside.
The \textit{no-hair} hypothesis that astrophysical black holes
are indeed described by the family of Kerr black holes
is tested in current and future observations \cite{Cardoso:2016ryw}.

So far all observations are in agreement with the Kerr hypothesis,
be it the motion of stars around the supermassive black hole
at the center of the Milky Way (2020 Nobel Prize),
the observation of gravitational waves from black hole mergers
(2017 Nobel Prize) or the observation of the shadow of the
supermassive black hole at the center of M87 (EHT collaboration).
However, it is expected that GR will be superseded by a new
gravitational theory, that will include quantum mechanics,
and that might as well explain (part of) the cosmological dark components,
dark matter and dark energy.
Overviews of alternative theories of gravity are found, for instance,
in \cite{Will:2005va,Faraoni:2010pgm,Berti:2015itd,CANTATA:2021mgk}.

A particularly attractive type of alternative theories 
of gravity are theories that contain higher curvature terms in the form of the Gauss-Bonnet 
invariant, as they arise, for instance, in string theories \cite{Zwiebach:1985uq,Gross:1986mw,Metsaev:1987zx}.
Since in four dimensions the Gauss-Bonnet term corresponds to a topological term,
that does not contribute to the field equations, this term has to be coupled
to another field in order to make its presence count. In string theory this field
is a scalar field, a so-called dilaton, that arises with a specific exponential coupling function
to the Gauss-Bonnet term in the low energy limit of string theory.
We will refer to these theoretically well motivated theories in the following as 
Einstein-dilaton-Gauss-Bonnet (EdGB) theories.
EdGB theories do not allow for GR black hole solutions.
Instead all EdGB black hole solutions carry dilatonic hair
\cite{Kanti:1995vq,Torii:1996yi,Guo:2008hf,Pani:2009wy,Pani:2011gy,Kleihaus:2011tg,Ayzenberg:2013wua,Ayzenberg:2014aka,Maselli:2015tta,Kleihaus:2014lba,Kleihaus:2015aje,Blazquez-Salcedo:2016enn,Cunha:2016wzk,Zhang:2017unx,Blazquez-Salcedo:2017txk,Konoplya:2019hml,Zinhailo:2019rwd}.

In recent years, other interesting coupling functions 
for the scalar field have been suggested
\cite{Sotiriou:2013qea,Sotiriou:2014pfa,Antoniou:2017acq,Doneva:2017bvd,Silva:2017uqg}.
In the following we will call such theories simply Einstein-scalar-Gauss-Bonnet (EsGB) theories.
Like the EdGB theories, the EsGB theories possess the attractive features
that they give rise to second order equations of motion,
and do not possess Ostrogradski instabilites and ghosts
\cite{Horndeski:1974wa,Charmousis:2011bf,Kobayashi:2011nu}.

By allowing for more general coupling functions of the scalar field
a new interesting phenomenon was observed:
curvature induced spontaneous scalarization of black holes
\cite{Antoniou:2017acq,Doneva:2017bvd,Silva:2017uqg,Antoniou:2017hxj,Blazquez-Salcedo:2018jnn,Doneva:2018rou,Minamitsuji:2018xde,Silva:2018qhn,Brihaye:2018grv,
Myung:2018jvi,Bakopoulos:2018nui,Doneva:2019vuh,Myung:2019wvb,Cunha:2019dwb,
Macedo:2019sem,Hod:2019pmb,Bakopoulos:2019tvc, Collodel:2019kkx,Bakopoulos:2020dfg,
Blazquez-Salcedo:2020rhf,Blazquez-Salcedo:2020caw}.
In that case an appropriate choice of coupling function allows the GR black holes to remain solutions of the
EsGB equations, while, at critical values of the coupling, GR black holes develop a tachyonic instability where new branches
of spontaneously scalarized black holes arise.
Moreover, in the case of rotation, there can exist two types of spontaneously scalarized black holes.
Those, that arise simply from the static black holes in the limit of slow rotation,
and those that arise only for fast rotation and are termed
spin induced spontaneously scalarized black holes
\cite{Dima:2020yac,Hod:2020jjy,Doneva:2020nbb,Zhang:2020pko,Doneva:2020kfv,Herdeiro:2020wei,Berti:2020kgk}.

This paper is organized as follows: In section II we briefly recall some properties of black holes in GR.
We discuss black holes in EdGB theories in section III and black holes in EsGB theories in section IV.
In both cases we will address first the static and then the rotating black holes, where in the
EsGB case we then differentiate
between those, %note, that depending on the sign of the coupling constant, there are two
%types of rapidly rotating black holes, the ordinary ones,
 that emerge continuously from the static limit,
and the spin induced ones. %, that are present only for fast rotation.
We end in section V with our conclusions.

\section{Black Holes in General Relativity}

Since our aim is to address deviations of the properties of black holes
in certain alternative theories of gravity from the properties of 
black holes in GR, we will start with a brief recap
of some basic properties of black holes that arise as solutions of the
Einstein field equations in vacuum
\begin{equation}
G_{\mu\nu} =  R_{\mu\nu} -\frac{1}{2} R\, g_{\mu\nu}  =0 \ ,
\end{equation}
with Einstein tensor $G_{\mu\nu}$, Ricci tensor $R_{\mu\nu}$, curvature scalar $R$
and metric tensor $g_{\mu\nu}$.

The asymptotically flat, static, spherically symmetric black hole solutions 
of these vacuum field equations are the Schwarzschild black holes.
The corresponding set of rotating black holes is the family of Kerr black holes.
For these vacuum black holes there is the well-known \textit{no-hair} theorem \cite{Chrusciel:2012jk},
stating that a Kerr black hole
is uniquely characterized in terms of only two global parameters:
the mass $M$ and the %electric charge and the %spin 
angular momentum $J$. For the static Schwarzschild black  hole
the angular momentum vanishes, so the mass is the only parameter.

Moreover, all the multipole moments of Kerr black holes
are given in terms of these two quantities, the mass $M$ and the 
angular momentum $J$
\cite{Geroch:1970cd,Hansen:1974zz,Thorne:1980ru}
\begin{equation}
M_l + i S_l = M \left( i \frac{J}{M} \right)^l \ ,
\end{equation}
with mass $M_0=M$, and angular momentum $S_1=J$,
and multipole number $l$.
The quadrupole moment $Q$ is then given by
$M_2 = Q  = - \frac{J^2}{M} $.

Kerr black holes are subject to a bound on their angular momentum,
\begin{equation}
j = \frac{J}{M^2} \le  1 \ ,
\end{equation}
the so-called Kerr bound, reached in the extremal limit,
when the two horizons of the Kerr black  holes coincide.
Solutions beyond the Kerr bound represent naked singularities.
The Kerr black  holes have been analyzed in many further
respects. Their shadow, for instance, was obtained first by Bardeen
\cite{Bardeen:1973tla} and recently revisited numerous times,
because of its significance for the EHT observations.

\section{Black Holes in Einstein-Dilaton-Gauss-Bonnet Theories}

We now turn to black holes in EdGB theories,
providing first the theoretical settings,
and presenting then the static and rotating solutions
and some of their properties.

\subsection{Theoretical Settings}

The effective action for EdGB and EsGB theories reads
%motivated by the low-energy heterotic string theory
\begin{eqnarray}  
S=\frac{1}{16 \pi}\int d^4x \sqrt{-g} \left[R - \frac{1}{2}
 \partial_\mu \phi \,\partial^\mu \phi - U(\phi)
 + F(\phi) R^2_{\rm GB}   \right],
\label{act}
\end{eqnarray} 
where $R$ is the curvature scalar,
$\phi$ is the scalar field, $F(\phi)$ is the coupling function, and 
$U(\phi)$ is the potential,
and
\begin{eqnarray} 
R^2_{\rm GB} = R_{\mu\nu\rho\sigma} R^{\mu\nu\rho\sigma}
- 4 R_{\mu\nu} R^{\mu\nu} + R^2 
\end{eqnarray} 
is the Gauss-Bonnet term. 

Variation of the action with respect to the metric and the scalar field leads to 
the Einstein equations and to the scalar field equation, respectively,
\begin{eqnarray}
G_{\mu\nu} & = & T_{\mu\nu} \ , 
\label{Einsteq}\\
\nabla^\mu \nabla_\mu \phi & + & \dot{F}(\phi) R^2_{\rm GB}-\dot{U}(\phi)=0 \ ,
\label{scleq}
\end{eqnarray}
where the dot denotes the derivative with respect to the scalar field $\phi$.
The stress-energy tensor in the gravitational field equation is an effective one,
since it contains not only the usual contributions from the scalar field,
but also contributions from the Gauss-Bonnet term.
It is given by the expression
\begin{equation}
T_{\mu\nu} =
-\frac{1}{4}g_{\mu\nu}\left(\partial_\rho \phi \partial^\rho \phi + 2 U(\phi) \right)
+\frac{1}{2} \partial_\mu \phi \partial_\nu \phi
-\frac{1}{2}\left(g_{\rho\mu}g_{\lambda\nu}+g_{\lambda\mu}g_{\rho\nu}\right)
\eta^{\kappa\lambda\alpha\beta}\tilde{R}^{\rho\gamma}_{\phantom{\rho\gamma}\alpha\beta}\nabla_\gamma \partial_\kappa F(\phi) \ ,
\label{tmunu}
\end{equation}
where
$\tilde{R}^{\rho\gamma}_{\phantom{\rho\gamma}\alpha\beta}=\eta^{\rho\gamma\sigma\tau}
R_{\sigma\tau\alpha\beta}$ and $\eta^{\rho\gamma\sigma\tau}= 
\epsilon^{\rho\gamma\sigma\tau}/\sqrt{-g}$.
As mentioned above, this resulting set of coupled field equations is of second order.
Also note, that in four spacetime dimensions the coupling of the Gauss-Bonnet term
to another field is really needed
in order to allow for solutions that differ from those of GR.

Since in this section we will discuss dilatonic black holes, we now specify the
coupling function to the dilatonic coupling function
\begin{equation}
\displaystyle  F(\phi)=\frac{{\alpha}}{4} e^{-{\gamma} \phi} \ ,
\end{equation}
where $\alpha$ is the Gauss-Bonnet coupling constant, and
$\gamma$ is the dilaton coupling constant with string theory value
$\gamma=1$.
For this coupling function $\dot{F}(\phi) \ne 0$ unless $\phi \to \infty$.
Therefore the dilaton field equation (\ref{scleq}) does not allow
a constant value of $\phi$ as a solution, if the Gauss-Bonnet term is 
non-vanishing, as it would be the case for a Schwarzschild black hole.
Consequently, the Schwarzschild black  hole cannot be a solution
of EdGB theory, and neither can the Kerr black hole:
all EdGB black hole solutions necessarily carry dilaton hair.

\subsection{Dilatonic Black Holes}

Static, spherically symmetric black hole solutions of EdGB theory
were first obtained by Kanti et al.~\cite{Kanti:1995vq}.
Because of symmetry one can choose the ansatz
\begin{eqnarray}
ds^2= - e^{2\Phi(r)}dt^2 + e^{2\Lambda(r)} dr^2 + r^2 (d\theta^2 + \sin^2\theta d\varphi^2 ) \ 
\label{met}
\end{eqnarray}  
for the metric, with two metric functions $\Phi(r)$ and $\Lambda(r)$,
that depend only on the radial coordinate, like the dilaton function $\phi(r)$.
While the EdGB black hole solutions have not been found in closed form,
numerical integration has yielded their domain of existence and their properties \cite{Kanti:1995vq}.

\begin{figure}[t!]
\begin{center}
%
%(a)\includegraphics[width=.45\textwidth, angle =0]{fig1a.eps}
%(b)\includegraphics[width=.45\textwidth, angle =0]{fig1b.eps}
%\\
\includegraphics[width=.33\textwidth, angle =-90]{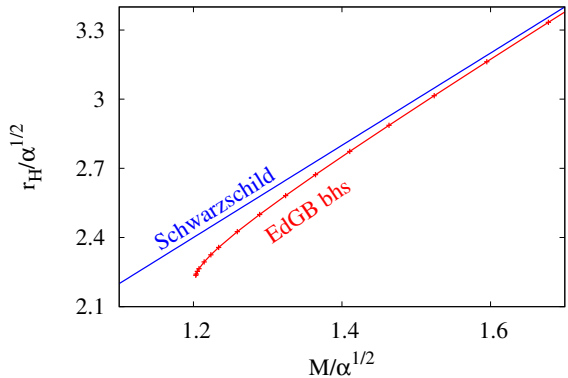}
\end{center}
\caption{Static spherically symmetric EdGB black holes ($\gamma=1$): scaled horizon radius $r_H/\sqrt{\alpha}$
vs scaled mass $M/\sqrt{\alpha}$.
For comparison the Schwarzschild solutions are also shown.}
\label{Fig1}
\end{figure}

We show in Fig.~\ref{Fig1} the scaled horizon radius versus the
scaled mass and compare with the Schwarzschild black hole.
Clearly, for large masses, the EdGB black holes approach the Schwarzschild black holes,
whereas for small masses the deviation from the Schwarzschild black holes becomes large.
Surprisingly, one finds a minimal value of the mass
for these EdGB black holes.
The reason can be found in the expansion of the functions at the horizon.
Here a square root appears
\begin{equation}
\sqrt{ 1-6 \frac{\alpha^2}{r_H^4} e^{2 \gamma \phi_H}} \ ,
\end{equation}
whose radicand vanishes at the minimal value of the mass.
We will refer to such solutions as critical solutions.
Depending on the value of the dilaton coupling constant $\gamma$,
a tiny second branch may exist, where the mass increases
again slightly until the horizon becomes singular \cite{Torii:1996yi,Guo:2008hf}.

The static EdGB black holes can be generalized to include rotation, either perturbatively
or by non-perturbative numerical calculations
\cite{Pani:2009wy,Pani:2011gy,Kleihaus:2011tg,Ayzenberg:2013wua,Ayzenberg:2014aka,Maselli:2015tta,Kleihaus:2014lba,Kleihaus:2015aje}.
The non-perturbative solutions can for instance be obtained with the
stationary axially symmetric line element \cite{Kleihaus:2011tg,Kleihaus:2014lba,Kleihaus:2015aje}
\begin{eqnarray}
\label{met2}
ds^2=- f dt^2 +  \frac{m}{f} \left( d r^2+ r^2d\theta^2 \right) 
           +  \frac{l}{f} r^2\sin^2\theta (d\varphi-\frac{\omega}{r} dt)^2 ,
\end{eqnarray}
with quasi-isotropic radial coordinate $r$.
The metric functions
$f$, $m$, $l$ and $\omega$ depend on $r$ and $\theta$ only, 
and the scalar field is also a function of $r$ and $\theta$ only, 
$\phi=\phi(r,\theta)$.

\begin{figure}[h!]
\begin{center}
(a)\includegraphics[width=.45\textwidth, angle =0]{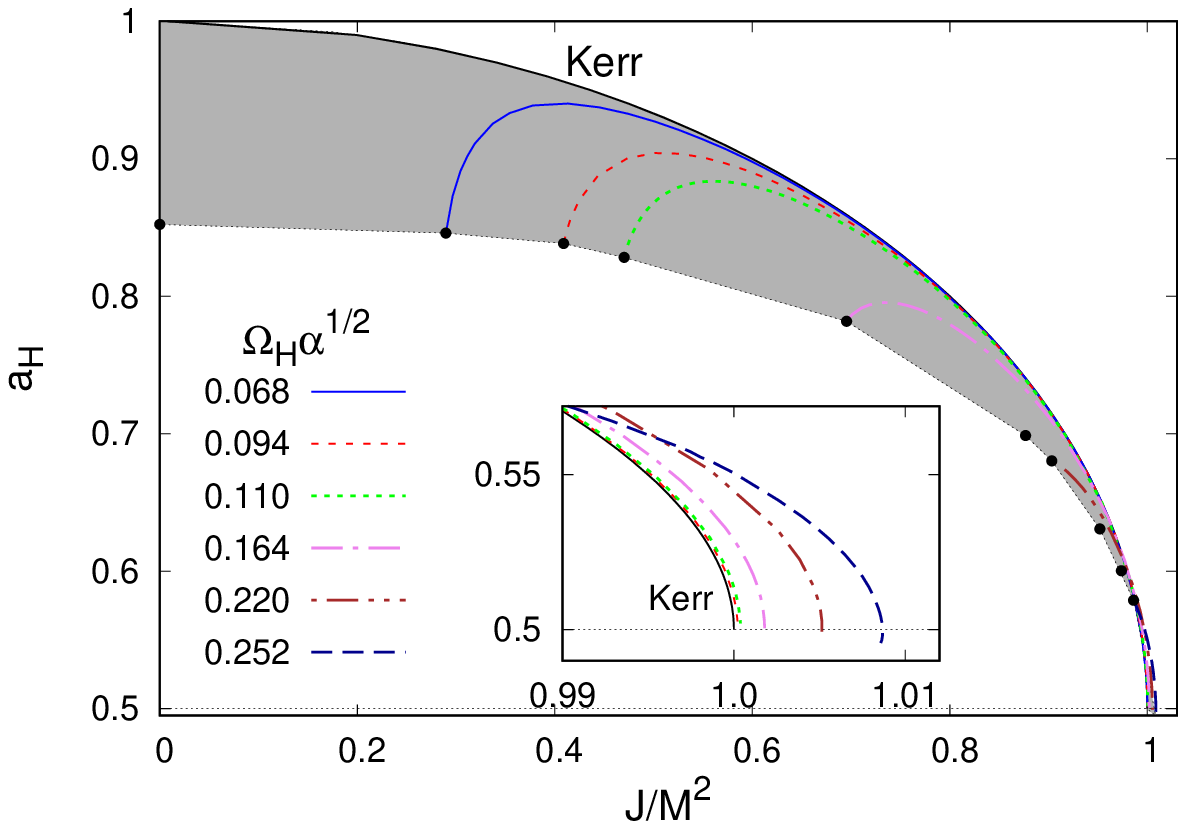}
(b)\includegraphics[width=.45\textwidth, angle =0]{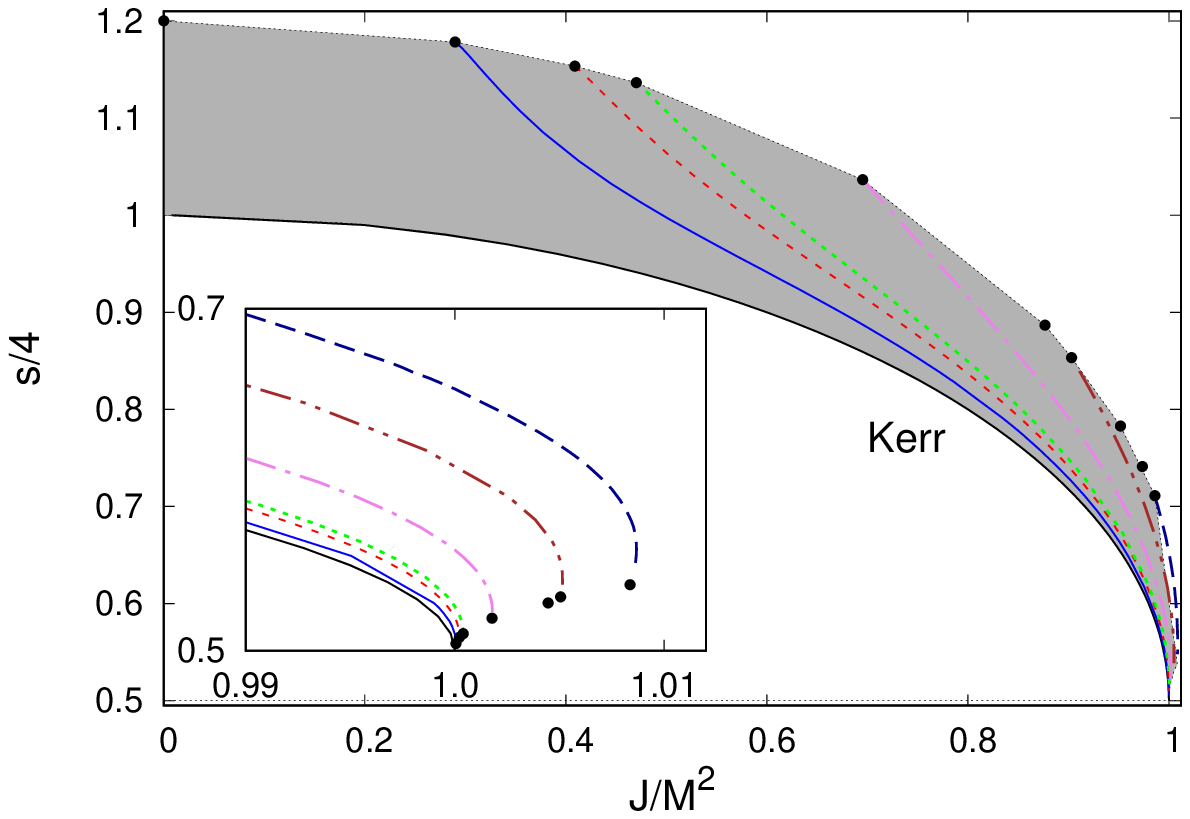}
%(a)\includegraphics[width=.45\textwidth, angle =0]{EGBd1b.eps}
%(b)\includegraphics[width=.45\textwidth, angle =0]{EGBd1a.eps}
%\\
%\includegraphics[width=.45\textwidth, angle =0]{Bamfig1bnn.eps}
\end{center}
\caption{Rotating EdGB black holes ($\gamma=1$): 
(a) scaled  horizon area $a_H=A_H/16 \pi r_H^2$ vs
scaled angular momentum $J/M^2$;
(b)  scaled entropy $s=S/16 \pi r_H^2$
vs scaled angular momentum $J/M^2$.
}
\label{Fig2}
\end{figure}

We present the domain of existence of these rotating EdGB black holes in Fig.~\ref{Fig2}.
In Fig.~\ref{Fig2}(a) the scaled horizon area $A_H/16 \pi r_H^2$ is shown versus the
scaled angular momentum $J/M^2$. For a fixed value of the coupling constant,
black  holes exist in the shaded region. The boundary of this region consists
of the static black  holes (left vertical boundary), the Kerr black holes (mostly upper boundary)
and the critical black holes (mostly lower boundary).
Very close to the Kerr bound $J/M^2=1$ these two boundaries cross and interchange.
The last boundary is only seen in the inset in the figure, and shows the extremal
black holes, which are not regular, however, in the EdGB case.
Clearly, in a small part of the domain of existence the Kerr bound is slightly exceeded
by almost extremal EdGB black holes. The curves inside the plot represent curves of
constant horizon angular velocity.

Fig.~\ref{Fig2}(b) shows the entropy of these black holes.
In GR black holes
possess an entropy that is simply a quarter of the event horizon area.
However, in the presence of a Gauss-Bonnet term, coupled to a scalar field,
the entropy of the EdGB black holes acquires an extra contribution \cite{Wald:1993nt}.
Then the total entropy
can be written in Wald's form 
as an integral over the event horizon
\begin{eqnarray}
\label{S-Noether} 
S=\frac{1}{4}\int_{\Sigma_{\rm H}} d^{2}x 
\sqrt{h}(1+ \frac{1}{2}\alpha e^{-\gamma \phi} \tilde R)  \ ,
\end{eqnarray} 
where $h$ is the determinant of the induced metric on a spatial cross section of the horizon 
and $\tilde R$ is the event horizon curvature. 
The figure shows, that the dilatonic black holes have larger entropy
than the Kerr black holes, while they have smaller horizon area.

Considering further properties of the rotating EdGB black holes
we note, that they can possess much larger quadrupole moments than Kerr black holes,
and their ISCOs and orbital frequencies can deviate appreciably from 
the respective Kerr values, as well.
Since their horizon area is smaller than for Kerr black  holes,
one might also expect considerable deviations for their shadow
as compared to the Kerr black  hole shadow. However,
these deviations turn out to be rather small \cite{Cunha:2016wzk}.
Also the x-ray reflection spectrum of accreting EdGB black  holes
shows only small deviations from the Kerr case
\cite{Zhang:2017unx}.

\subsection{Linear Mode Analysis: Quasi-Normal Modes}

To investigate stability of black holes under small perturbations, 
a linear mode analysis can be performed. 
Here we will directly address the formalism for
quasi-normal modes. Since in gravity small perturbations
will typically lead to the emission  of gravitational waves
the frequencies that are found in perturbation theory
also contain an imaginary part, which explains the terminology.
From an observational point of view, such quasi-normal modes
will appear in the ringdown spectra of black holes after merger.
This makes their study most relevant in connection with
current and future gravitational wave observations.

While we refrain from a full derivation of these quasi-normal 
modes and refer  to the literature 
\cite{Kokkotas:1999bd,Nollert:1999ji,Rezzolla:2003ua,Berti:2009kk,Konoplya:2011qq}, 
we briefly recall some of the relevant aspects.
To this end we consider lowest order perturbation theory 
 in the metric
\begin{equation}
g_{\mu\nu} = g_{\mu\nu}^{(0)}(r) + \epsilon h_{\mu\nu}(t,r,\theta,\varphi)
\end{equation}
and in the scalar field
\begin{equation}
\phi = \phi_0(r) + \epsilon \delta \phi(t,r,\theta,\varphi) \ ,
\end{equation}
where $g_{\mu\nu}^{(0)}$ and $\phi_0$ are the metric 
and the scalar field of the background black hole, respectively,
and $h_{\mu\nu}$ and $\delta \phi$ are the perturbations.
$\epsilon$ is the small perturbation parameter.

Symmetry allows for a decomposition of the perturbations
into even-parity and odd-parity perturbations.
The scalar field has even parity, therefore it decouples in the case
of odd-parity perturbations, which are therefore
pure spacetime modes.
The even-parity modes are also called polar modes,
while the odd-parity modes are also termed axial modes.
Besides the decomposition with respect to parity,
we can also make a multipolar decomposition of the modes,
characterized by the angular parameter $l$.

There are quasi-normal modes for all values of the
angular parameter $l$.
Because of the scalar field also modes with 
angular parameter $l=0$ and $l=1$ arise,
corresponding to radial modes (monopole modes)
and dipole modes.
For $l=2$ quadrupole modes arise, which are also present
in GR. But because of the scalar field, there will now be
two types of such modes: $l=2$ modes dominated
by the scalar field, which in the limit of vanishing
Gauss-Bonnet coupling would correspond to 
modes of the scalar field in the background of a Schwarzschild black hole,
and $l=2$ modes dominated by the gravitational field,
which would correspond to the lowest Schwarzschild
quadrupolar modes.
We will refer to the first set of modes as scalar-led modes,
and to the second set as grav-led modes in the following.

The time dependence of the modes is factored out by an
exponential 
\begin{equation} 
\exp{( i \omega t)} = \exp{( i (\omega_R + i \omega_I)t)}
= \exp{(i \omega_R t - \omega_I t )} \ ,
\end{equation} 
which shows that the real part $\omega_R$ is the frequency
and the imaginary part $\omega_I$ is the inverse damping time
if $\omega_I>0$, otherwise, for $\omega_I<0$, it signals an instability.
(Note, that the overall sign choice 
in the exponent is only convention.) 
For a given parity and angular parameter $l$
the complex frequency $\omega$ is obtained
by solving the respective resulting system of coupled 
differential equations, subject to proper boundary conditions.
At the black hole horizon the wave must be purely ingoing,
and at infinity purely outgoing.

\begin{figure}[h!]
\begin{center}
(a)\includegraphics[width=.33\textwidth, angle =-90]{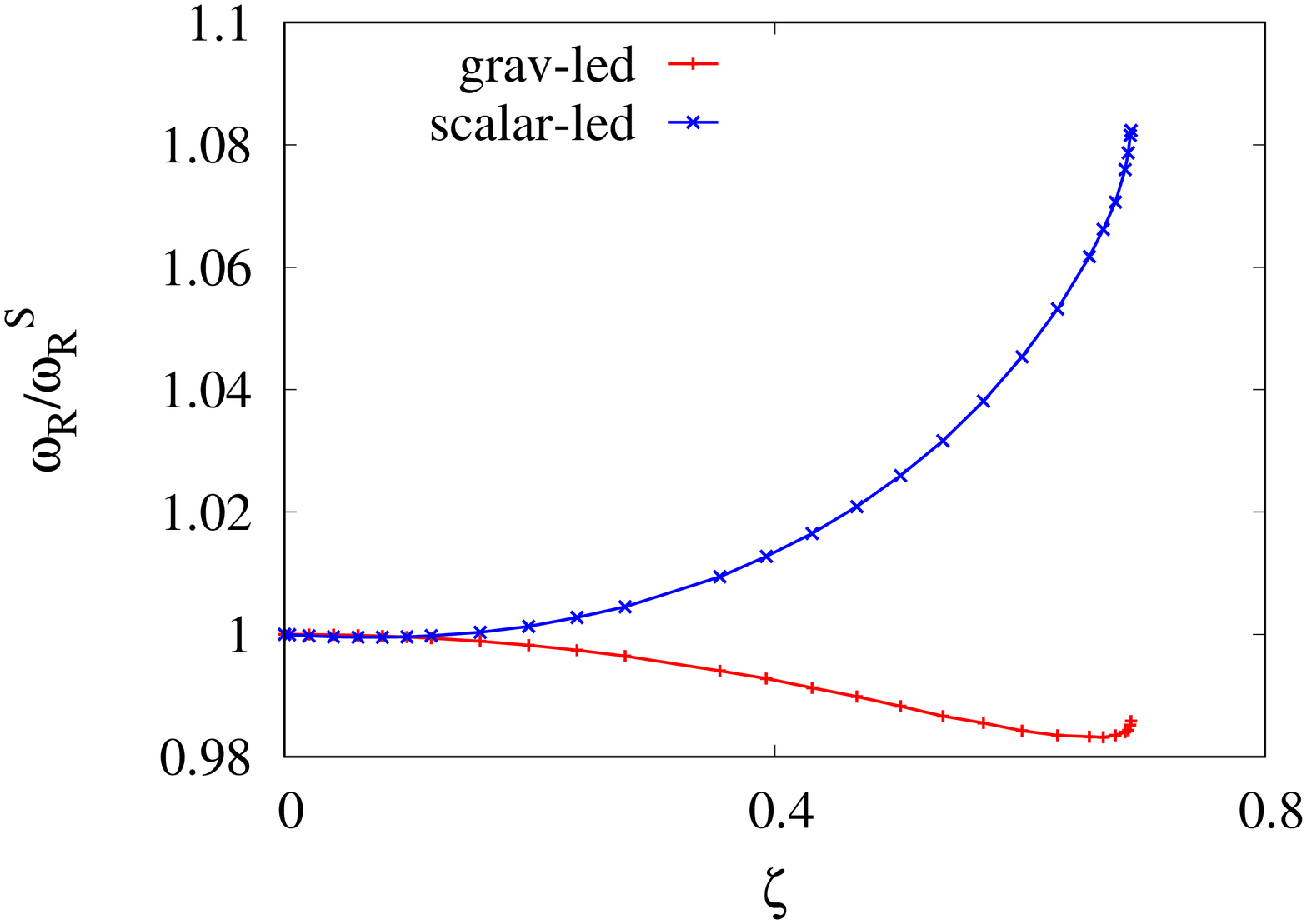}
(b)\includegraphics[width=.33\textwidth, angle =-90]{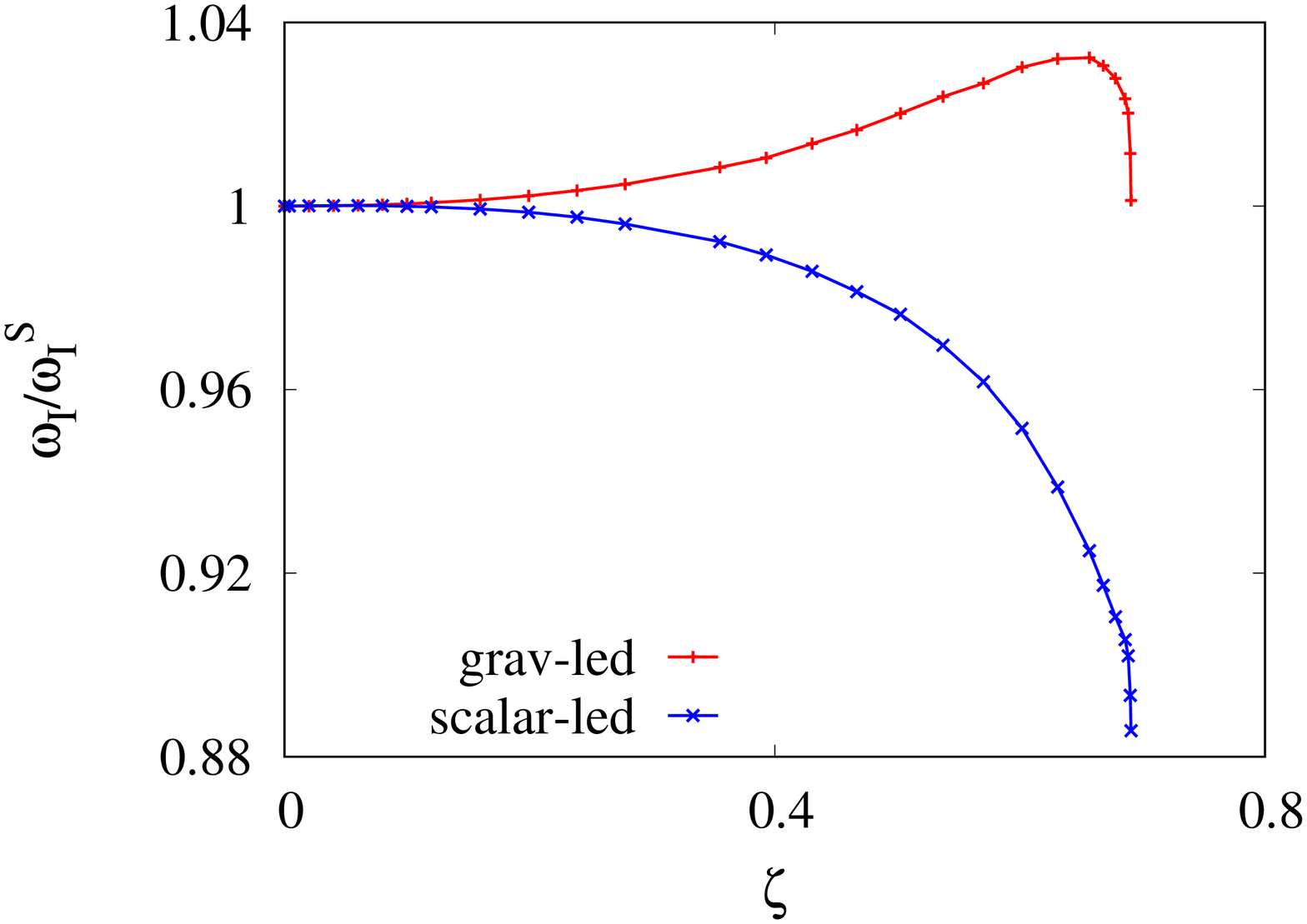}
%\\
%\includegraphics[width=.45\textwidth, angle =0]{Bamfig1bnn.eps}
\end{center}
\caption{Quasi-normal polar $l=2$ modes of static EdGB black holes ($\gamma=1$): 
(a) scaled frequency $\omega_R/\omega_R^S$
vs scaled Gauss-Bonnet coupling constant $\zeta=\alpha/M^2$;
(b) scaled inverse damping time $\omega_I/\omega_I^S$
vs scaled Gauss-Bonnet coupling constant $\zeta=\alpha/M^2$.
}
\label{Fig3}
\end{figure}

As an interesting example we show in Fig.~\ref{Fig3}
the quasi-normal polar $l=2$ modes for the static
EdGB black holes, normalized to the respective
Schwarzschild values for vanishing coupling constant.
Fig.~\ref{Fig3}(a) shows the real part of $\omega$,
and Fig.~\ref{Fig3}(b) the imaginary part
versus the scaled Gauss-Bonnet coupling constant $\zeta=\alpha/M^2$.
We note a distinctly different behavior for the
grav-led and the scalar-led modes. 
We also note, that the presence of the scalar field in the
polar modes breaks isospectrality of the modes,
i.e., the degeneracy of the axial and polar $l=2$ 
gravitational modes in the Schwarzschild case.

\section{Black Holes in Einstein-Scalar-Gauss-Bonnet Theories}

Whereas EdGB theories are already considerably constrained from observations,
this is much less the case for EsGB theories with more general coupling functions.
We now turn to the black holes in these theories and focus on
coupling functions which allow for spontaneous scalarization.

\subsection{Curvature Induced Spontaneous Scalarization}

The phenomenon of spontaneous scalarization was discovered for
neutron stars in scalar-tensor theories \cite{Damour:1993hw},
where GR neutron stars can develop a scalar field
when the solutions become sufficiently compact.
Here the trigger for the scalarization is the highly compact
matter. Therefore the spontaneous scalarization  is referred
to as matter induced spontaneous scalarization.
The absence of matter for Schwarzschild and Kerr black holes
therefore precludes this phenomenon for these black  holes.

Only a few years ago it was realized that spontaneous
scalarization can also be curvature induced and therefore
arise for black holes in EsGB theories
\cite{Antoniou:2017acq,Doneva:2017bvd,Silva:2017uqg}.
In order to allow for such spontaneous scalarization
the coupling function should possess certain properties.
First of all, the GR black hole solutions should remain
solutions of the theory. This is of course the case, when
the Gauss-Bonnet term does not contribute in the field equations.
So if we choose a coupling function $F(\phi)$ such that
\begin{equation}
\dot F (\phi)=0 \ \ \ \ \text{for} \ \ \ \  \phi=0 \ ,
\end{equation}
then the source term in the scalar field equation
\begin{equation}
\nabla^\mu \nabla_\mu \phi + \dot F (\phi) R^2_{\rm GB}=0  \ 
\label{scalareq}
\end{equation}
vanishes for $\phi=0$, and $\phi=0$ is a solution.
Note, that we have assumed a vanishing scalar field potential $U(\phi)$
at the moment.
The Einstein equations then also receive no contribution from
the Gauss-Bonnet term, and therefore the GR solutions remain 
solutions of such EsGB theories.
However, the GR solutions are not the only black hole solutions
in certain parameter ranges, that depend on the coupling function.
Here black holes with scalar hair arise, and this hair is curvature induced.

To understand this mechanism, we consider the Gauss-Bonnet term
for the metric of a Schwarzschild black hole
\begin{equation}
R^2_{\rm GB} = \frac{48 M^2}{r^6} \ ,
\end{equation}
which is solely coming from the Kretschmann scalar.
Clearly, this curvature term can become rather big.
We now choose the simple coupling function
\begin{equation}
F(\phi) = \eta \frac{\phi^2}{2} \ , \ \ \
 \dot F = \eta \phi \ .
 \label{quadratic}
 \end{equation}
When we insert this into the scalar field equation,
we see, that we can identify an effective mass squared $m^2_{\rm eff}$
in this equation
\begin{equation}
m^2_{\rm eff} = - \eta R^2_{\rm GB} < 0 \ , \ \ \ \text{if} \ \eta> 0 \ ,
\label{effm}
\end{equation}
and this effective mass squared is negative, i.e., tachyonic, for positive
coupling constant $\eta$.
Therefore the Gauss-Bonnet curvature term
triggers a tachyonic instability of the Schwarzschild solution,
when its contribution is strong enough,
and a branch of scalarized black holes bifurcates from
the Schwarzschild solution.

\subsection{Static Black Holes}

\begin{figure}[h!]
\begin{center}
(a)\includegraphics[width=.33\textwidth, angle =-90]{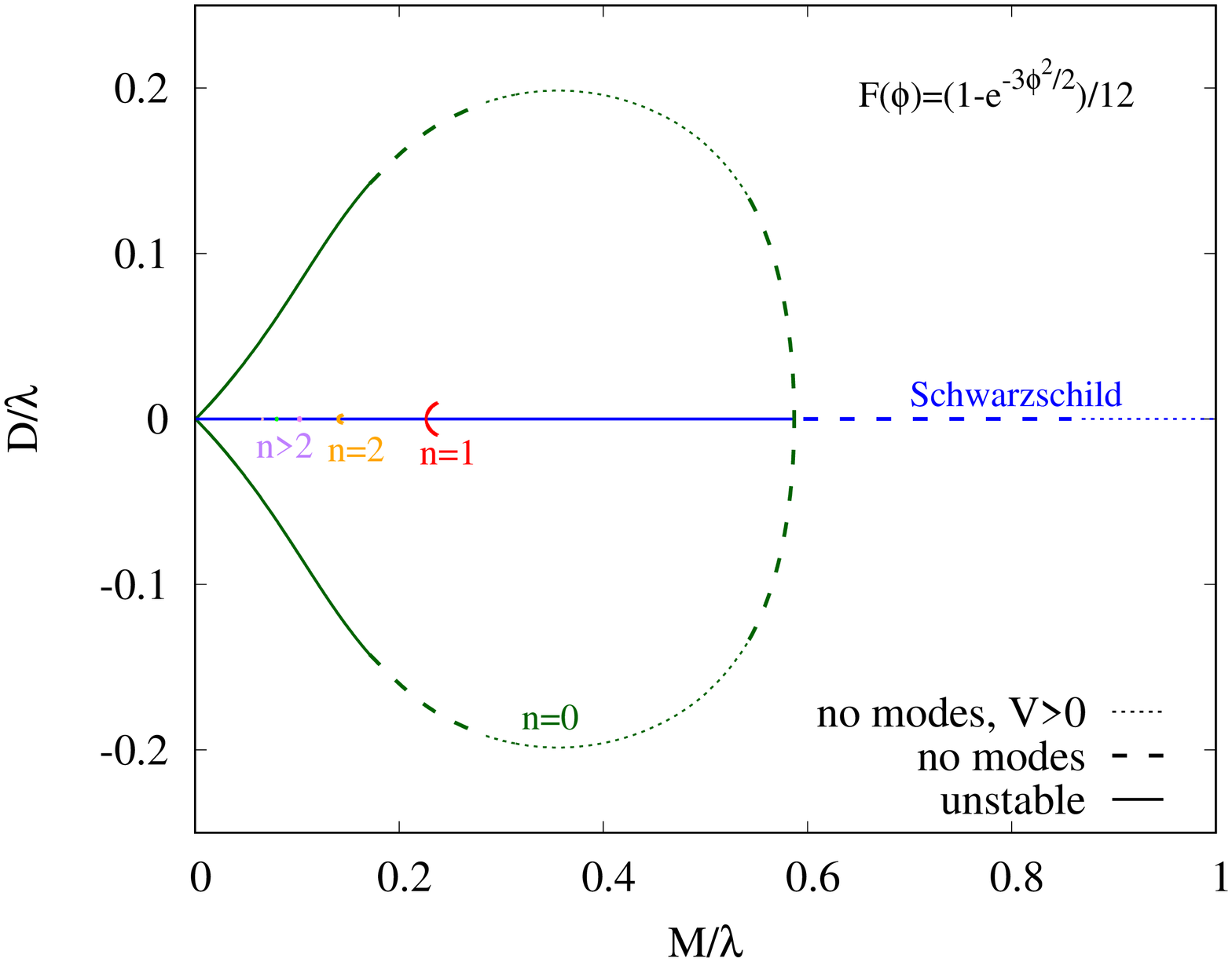}
(b)\includegraphics[width=.33\textwidth, angle =-90]{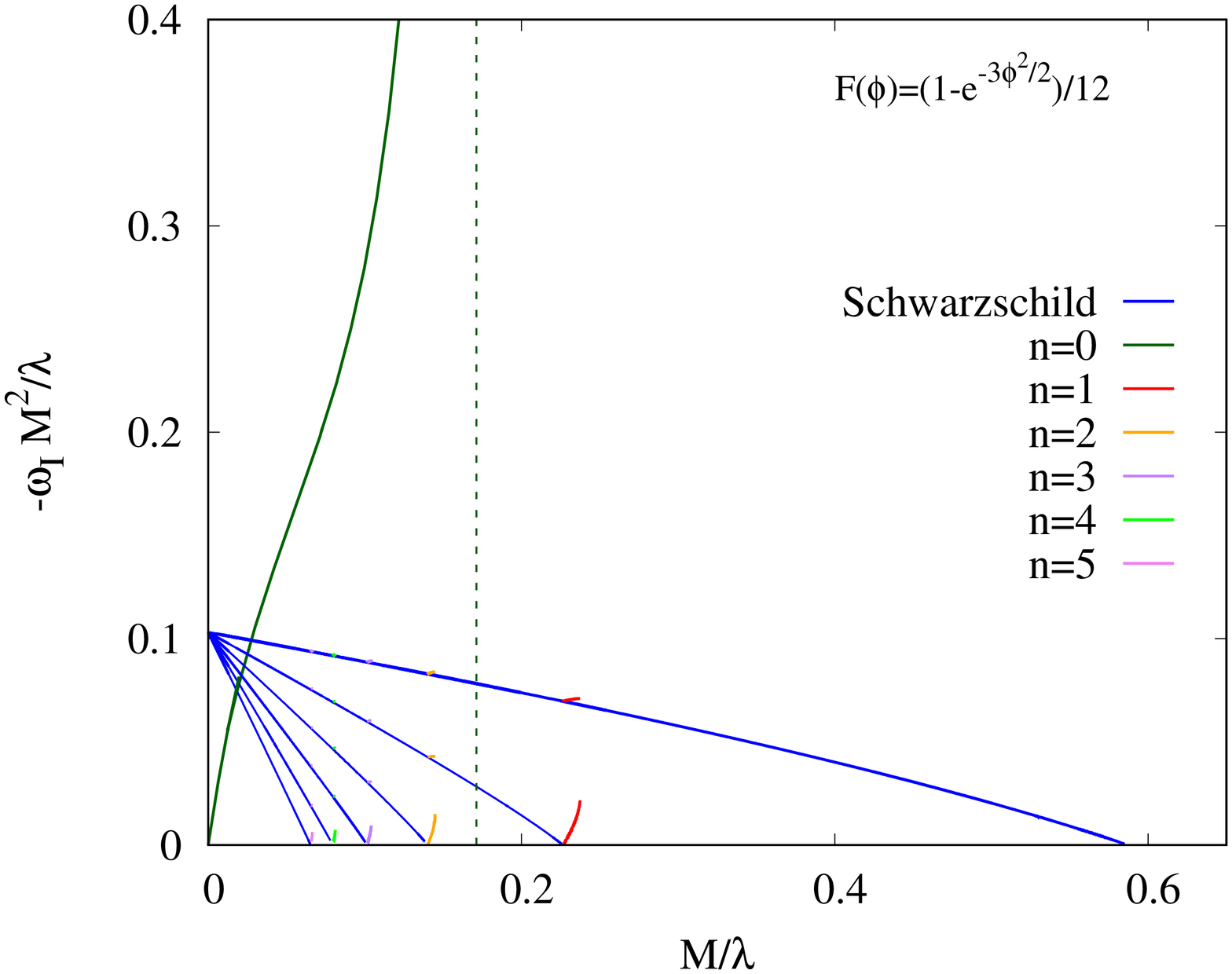}
%\\
%\includegraphics[width=.45\textwidth, angle =0]{Bamfig1bnn.eps}
\end{center}
\caption{Static EsGB black holes: (a) scaled scalar charge $D/\lambda$
vs scaled mass $M/\lambda$ for the fundamental ($n=0$)
and radially excited ($n>0$) solutions;
(b) scaled imaginary frequency $\omega_I M^2/\lambda$ 
of the unstable radial modes
vs scaled mass $M/\lambda$ for the fundamental ($n=0$)
and radially excited ($n>0$) solutions.
The Schwarzschild solution and its unstable modes
are also shown for comparison.
}
\label{Fig4}
\end{figure}

We now consider the coupling function \cite{Doneva:2017bvd}
\begin{equation}
F(\phi) =  \frac{ \lambda^2}{12} \left(1- e^{-3\phi^2/2}\right) \ ,
\label{exponential}
\end{equation}
which for small $\phi$ becomes simply a quadratic coupling,
$F(\phi)=  (\lambda^2/8) \phi^2$.
The tachyonic instability then arises at $M/\lambda=0.587$,
where a branch of scalarized black holes emerges.
Since this is the first bifurcation, we refer to this branch as the
fundamental or $n=0$ branch.
But this branch is not the only one, and at smaller values of 
$M/\lambda$ further branches arise. These are radially excited branches,
where the scalar field function possesses $n$ nodes.
Thus on the first excited branch ($n=1$), which arises at $M/\lambda=0.226$,
the scalar field has one node, on the second excited branch
($n=2$: $M/\lambda=0.140$) it has two nodes, etc.
When one follows these scalarized branches, as shown in
Fig.~\ref{Fig4}(a), one notes, that only the fundamental
branch extends from the bifurcation all the way to 
vanishing mass ($M=0$). The excited branches 
all have finite extent, and they are the shorter
the higher $n$.
Note, that in the figure the scaled scalar charge $D/\lambda$,
which is read off from the asymptotic behavior of the scalar field
($\phi \sim D/r$) is shown versus the scaled mass $M/\lambda$.
To highlight the bifurcations, also the Schwarzschild black hole
is shown, which of course carries no scalar charge.

Let us now consider the stability of these solutions,
in particular, we would like to know,
whether the fundamental scalarized solution is stable,
when it emerges from the Schwarzschild solution,
since the Schwarzschild solution has to become unstable
to develop scalar hair (tachyonic instability).
A first indication of stability is easily obtained by evaluating the
entropy of the fundamental scalarized solution
and comparing it to the entropy of the Schwarzschild solution \cite{Doneva:2017bvd}.
This shows, that the $n=0$ solution has higher entropy,
and should therefore be (thermodynamically) preferred.

The next step is to consider radial ($l=0$) perturbations \cite{Blazquez-Salcedo:2018jnn},
which are polar perturbations involving the scalar field.
When the Schr\"odinger-like master equation for the eigenvalue $\omega$
is solved for the Schwarzschild background, 
a zero mode is found precisely at the first bifurcation point.
As $M/\lambda$ is further decreased, this
zero modes turns into a negative mode,
as seen in Fig.~\ref{Fig4}(b).
In fact, at each bifurcation, where a new
branch of  radially excited scalarized black holes arises,
another zero mode of the Schwarzschild solution appears,
that turns into another unstable mode for smaller values of $M/\lambda$.

When we solve the Schr\"odinger-like master equation for the radial
perturbations in the background of the fundamental scalarized black hole
solutions, however, no radially unstable modes are found
in the region from the bifurcation up to a critical value of $M/\lambda$,
that is marked by the vertical dashed line in  Fig.~\ref{Fig4}(b).
Here the perturbation equation loses hyperbolicity
and the employed formalism breaks down. 
Let us denote this point by S1 for later reference
and turn to the radially excited branches.

Fig.~\ref{Fig4}(b) also shows the radially unstable modes
for the excited branches.
Since these branches emerge from the Schwarzschild black hole
at their respective bifurcation point, continuity at this bifurcation point demands,
that the unstable modes of the radially excited black holes also bifurcate there from 
the Schwarzschild zero and unstable mode(s).
So for the $n=1$ solution we observe two unstable modes,
one starting at the bifurcation point at the zero mode
and one starting at the first unstable Schwarzschild mode.
For the $n=2$ solution we then have three unstable modes,
etc.

While it is expected that radially excited solution are unstable,
it would be nice, if the fundamental branch really were stable.
So far we have only considered the $l=0$ modes. Therefore
we now turn to modes with higher $l$.
These were analyzed in \cite{Blazquez-Salcedo:2020rhf,Blazquez-Salcedo:2020caw}.
Since axial modes do not involve perturbations of the scalar field,
they start with the quadrupolar case $l=2$.
The analysis shows, that no further instability arises here,
however, hyperbolicity of the equations is lost slightly earlier
than in the radial case. Denoting this second point of
loss of hyperbolicity by S2, we note, that there is no
axial mode  instability between the bifurcation point and S2
for the fundamental branch.
Similarly, when the polar modes with $l=1$ (dipole)
and $l=2$ quadrupole are considered, no further
instability is encountered.
Thus we conclude, that the fundamental branch is mode
stable in the region between its bifurcation point
and the point S2.

\begin{figure}[h!]
\begin{center}
\mbox{
(a)\includegraphics[width=.33\textwidth, angle =270]{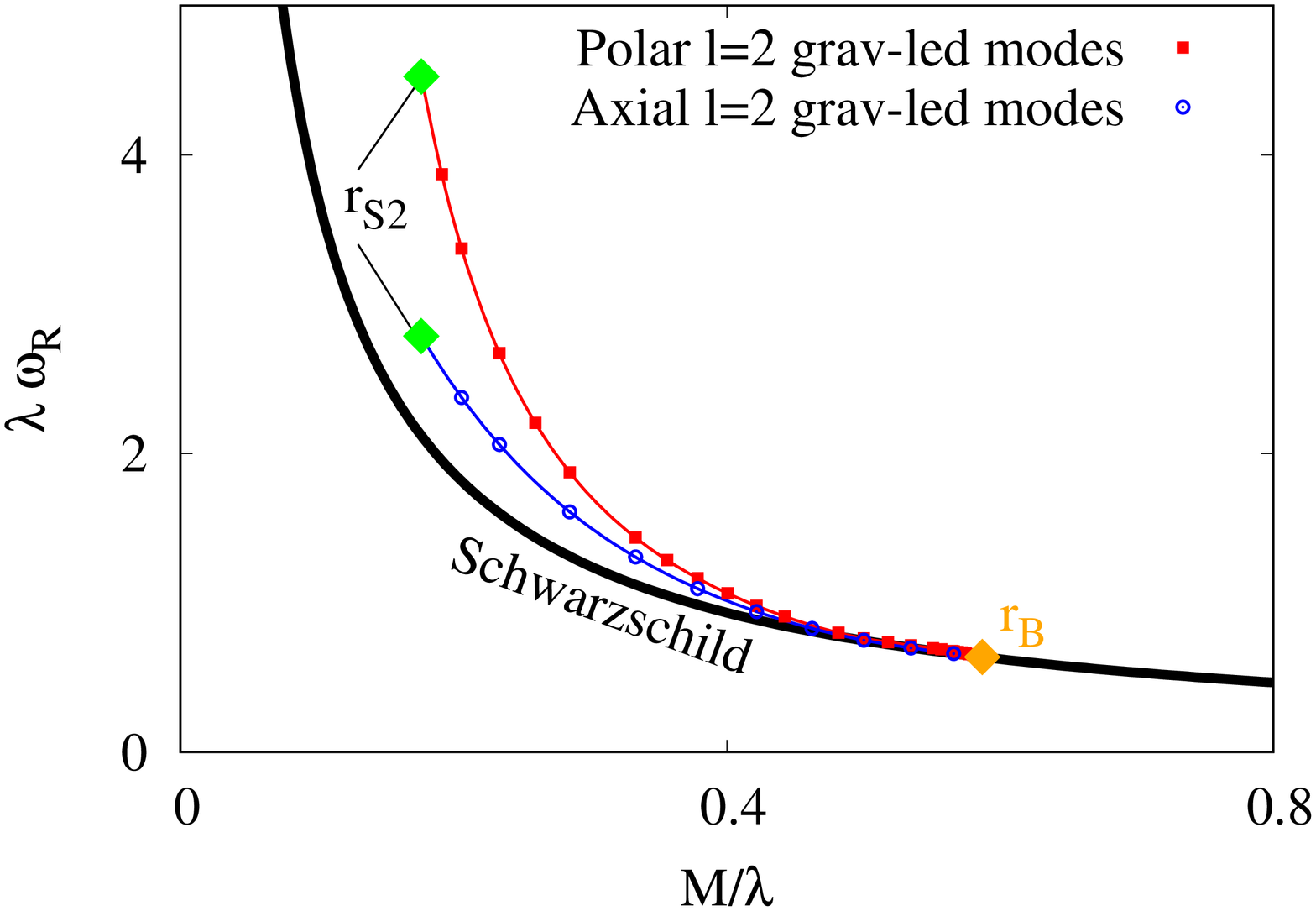}
(b)\includegraphics[width=.33\textwidth, angle =270]{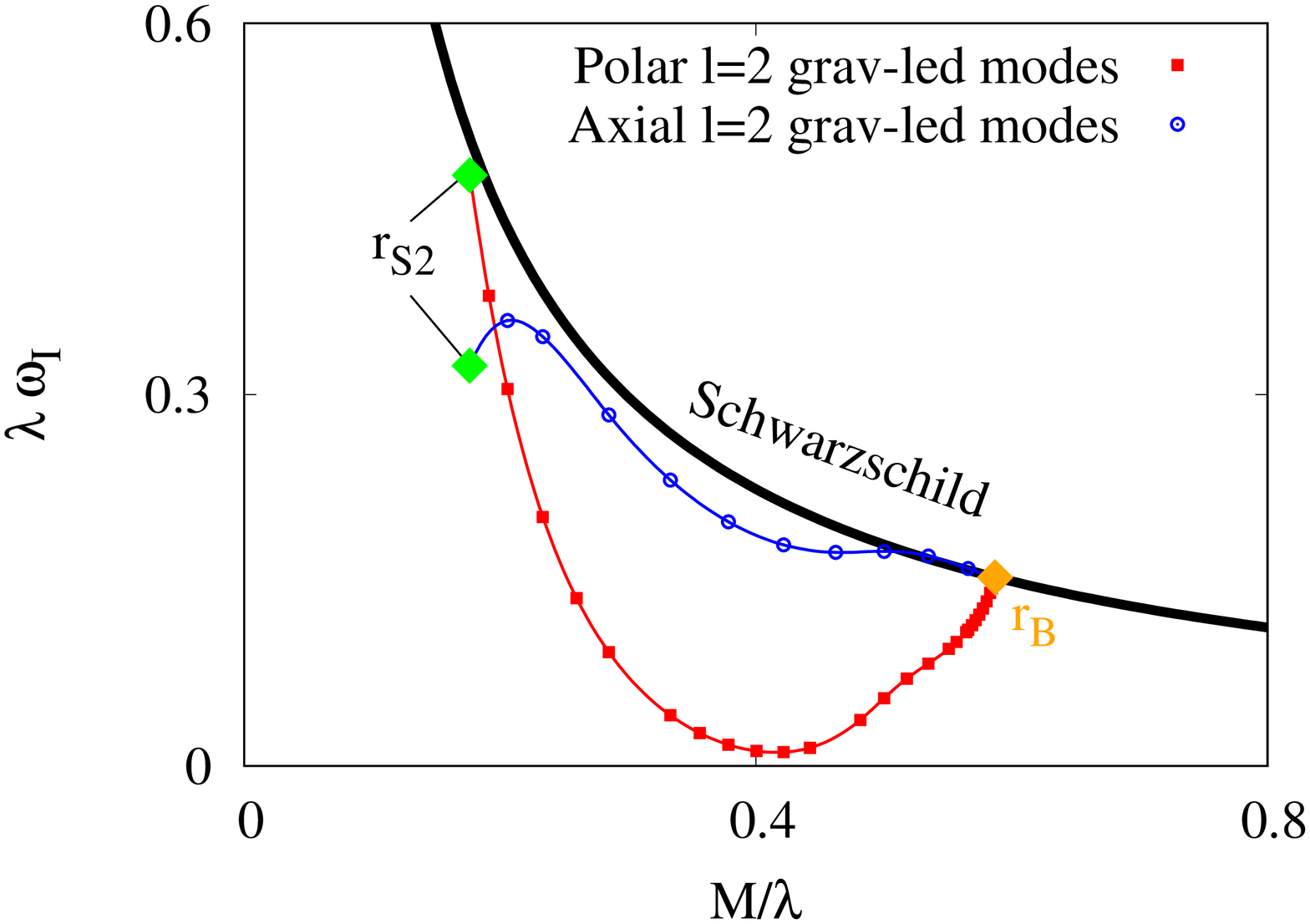}
}
%\\
%\includegraphics[width=.45\textwidth, angle =0]{Bamfig1bnn.eps}
\end{center}
\caption{Polar and axial $l=2$ grav-led modes of
static EsGB and Schwarzschild black holes: (a) scaled real part $\omega_R/\lambda$
vs scaled mass $M/\lambda$ for the fundamental ($n=0$) solution;
(b) scaled imaginary part $\omega_I/\lambda$ 
vs scaled mass $M/\lambda$ for the fundamental ($n=0$).
Note the isospectrality of the Schwarzschild modes.
}
\label{Fig5}
\end{figure}

While there are no new unstable modes, there are of course
numerous stable modes, where the imaginary part of the
eigenvalue is positive and corresponds to an inverse damping time.
As an example we exhibit the lowest such axial and polar grav-led $l=2$ (quadrupole)
modes in Fig.~\ref{Fig5}.
The figure nicely shows the degeneracy of these modes for the Schwarzschild case,
i.e., the isospectrality of the Schwarzschild modes.
In contrast, for the fundamental scalarized black holes
isospectrality is broken and the axial and polar modes generically differ.

A similar analysis can, in principle, also be performed for other coupling functions.
The simplest coupling function is of course the quadratic one, Eq.~(\ref{quadratic}).
Here already the entropy indicates instability of the fundamental branch
of scalarized black holes, and a radial mode analysis shows,
that the scalarized static spherically symmetric black  holes are indeed all
unstable including the fundamental branch \cite{Blazquez-Salcedo:2018jnn}.
Moreover, this branch is rather short, and oriented toward
larger values of $M/\lambda$ unlike the fundamental branch for
the exponential coupling function, Eq.~(\ref{exponential}).
Obviously, stability and length depend significantly on the
coupling function. Including higher order terms
in the coupling function with an appropriate sign can stabilize the solutions,
as demonstrated in \cite{Silva:2018qhn}.
Another way to stabilize the solutions is to allow for an appropriate self-interaction potential
$U(\phi)$ of the scalar field, as shown in \cite{Macedo:2019sem}
for a quartic self-interaction.

\subsection{Rotating Black Holes}

With applications to astrophysics in mind, one has to include rotation of the black holes,
and thus consider the phenomenon of curvature induced scalarization in the presence of
rotation. Here the GR solution is of course the Kerr black  hole.
Therefore we have to inspect  the source term in the scalar field equation,
Eq.~\ref{scalareq}), i.e., the Gauss-Bonnet term for a Kerr black hole
\begin{equation}
R^2_{\rm GB} = \frac{48 M^2}{(r^2 + \chi^2)^6} 
\left( r^6 - 15 r^4 \chi^2 + 15 r^2 \chi^4 - \chi^6 \right)
\ , \ \ \ \chi=a \cos \theta \ ,
\label{KerrGB}
\end{equation}
where $a$ is the usual Kerr specific angular momentum.
Recalling Eq.~(\ref{effm}) for the effective mass (with positive 
coupling constant $\eta$)
and inserting the above expression for the Gauss-Bonnet term,
we conjecture that the presence of the new terms that depend on
the angular momentum suppresses the scalarization for large rotation,
since the source term becomes weaker in (part of) the region with
large curvature.
%$$ m^2 = - \eta R^2_{\rm GB} < 0 $$ %\ , \ \ \ \text{if} \ \eta> 0$$

\begin{figure}[h!]
\begin{center}
(a)\includegraphics[width=.45\textwidth, angle =0]{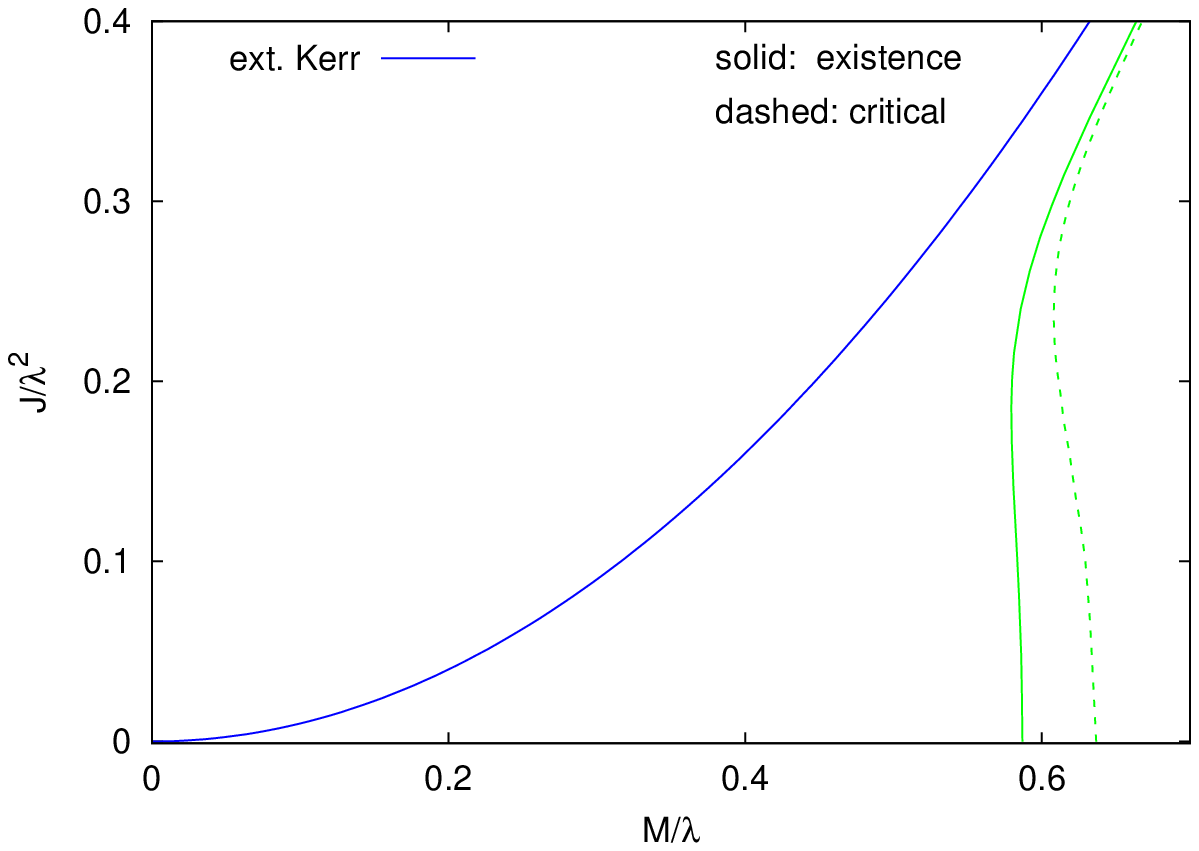}
(b)\includegraphics[width=.45\textwidth, angle =0]{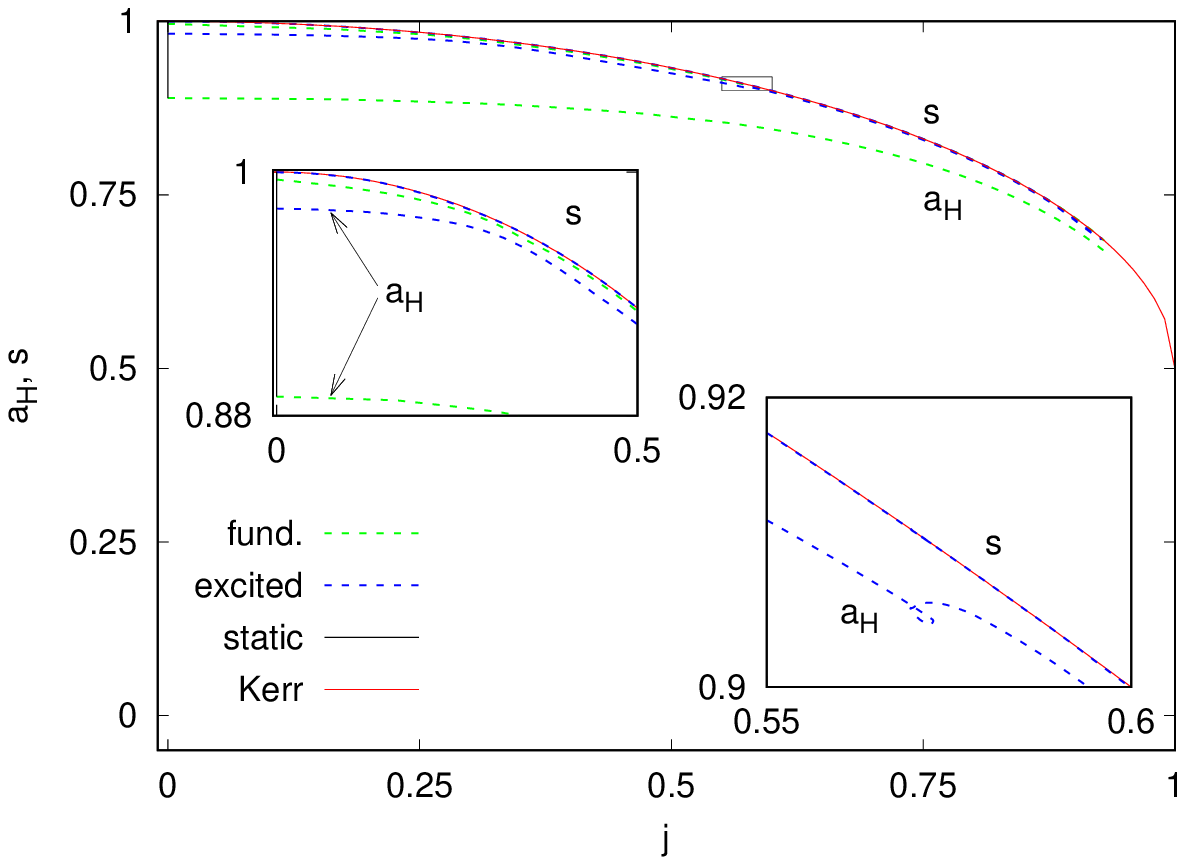}
%(a)\includegraphics[width=.45\textwidth, angle =0]{EsGBBufig1b.eps}
%(b)\includegraphics[width=.45\textwidth, angle =0]{EsGBBufigxe3bn.eps}
\end{center}
\caption{Rotating EsGB black holes $(\eta>0)$: (a) 
scaled angular momentum $J/\lambda^2$ vs scaled mass $M/\lambda$
for the existence line and the critical line, and also for the extremal Kerr black holes,
(b) scaled area $a_H=A_H/16 \pi r_H^2$
and scaled entropy $s=S/4\pi M^2$
vs scaled angular momentum $j=J/M^2$
for the fundamental and first radially excited black holes.
}
\label{Fig6}
\end{figure}

We begin the discussion of the rotating black hole solutions
and their properties by considering the quadratic coupling function, Eq.~(\ref{quadratic}),
constructed in \cite{Collodel:2019kkx}. We exhibit the domain of
existence of the fundamental scalarized branch in Fig.~\ref{Fig6}(a),
where the scaled angular momentum $J/\lambda^2$ is shown versus
the scaled mass $M/\lambda$, and we have introduced the coupling
constant $\eta=\lambda^2/8$, while keeping a vanishing 
self-interaction potential $U(\phi)=0$.
The figure contains three curves showing the extremal Kerr solutions,
the existence line for the scalarized black holes and the critical line
for the scalarized black holes (from left to right).
The existence line marks the onset of spontaneous scalarization,
while the critical line shows where the scalarized black holes
cease to exist.
Whereas the domain of Kerr black holes is the whole area below
the extremal curve, the domain of scalarized black holes
is only the small band between the existence line and the
critical line. As conjectured, the band becomes thinner
when the angular momentum is increased, i.e.,
angular momentum indeed suppresses the scalarization.

In Fig.~\ref{Fig6}(b) we show the scaled area $a_H=A_H/16 \pi r_H^2$
versus the scaled angular momentum $j=J/M^2$
for the fundamental solution and for the $n=1$ radial excitation.
The scaled entropy $s=S/4\pi M^2$ is also shown.
In this representation the Kerr black holes form the upper 
limiting curve, for which area and entropy agree.
As in the static case with quadratic coupling the entropy of the rotating fundamental black holes
is smaller than the entropy of the Kerr black holes.
Thus the instability persists for the fundamental scalarized black holes
with a quadratic coupling function and no self-interaction
when rotation is included.

In \cite{Cunha:2019dwb} the rotating fundamental scalarized black holes
were obtained for the exponential coupling function, Eq.~(\ref{exponential}).
Here the static fundamental branch is much larger 
and (at least to a large extent) also stable.
Starting from this large interval of static solutions, the
domain of existence is therefore much larger for the 
rotating solutions for this coupling function.
However, for fast rotation, the domain narrows again strongly,
leaving only a small band of rapidly rotating
scalarized black holes.
We note, that an interesting consequence of the broad range of 
slowly rotating black  holes is the possibility
to obtain a limit on the Gauss-Bonnet coupling constant,
by comparing the EsGB black hole shadow with observations
\cite{Cunha:2019dwb}.

Let us  now consider a final twist
concerning curvature induced rotating scalarized black holes.
To that end we return to the Gauss-Bonnet term
evaluated for a Kerr black  hole, Eq.~(\ref{KerrGB}).
Above we have noticed the strong suppressive effect of fast rotation
for spontaneous scalarization.
Now we would like to make use of this effect in a new constructive way.
As noticed in \cite{Dima:2020yac}
and further elaborated on in \cite{Hod:2020jjy,Doneva:2020nbb,Zhang:2020pko,Doneva:2020kfv},
a new way of inducing the tachyonic instability in the scalar field equation
is obtained for sufficiently fast rotation,
when a negative coupling constant $\eta<0$ is chosen 
in the coupling function $F(\phi)$.
Therefore this type of spontaneous scalarization is termed
spin induced spontaneous scalarization.
Its onset happens at a Kerr rotation parameter of $j=0.5$.

\begin{figure}[h!]
\begin{center}
(a)\includegraphics[width=.45\textwidth, angle =0]{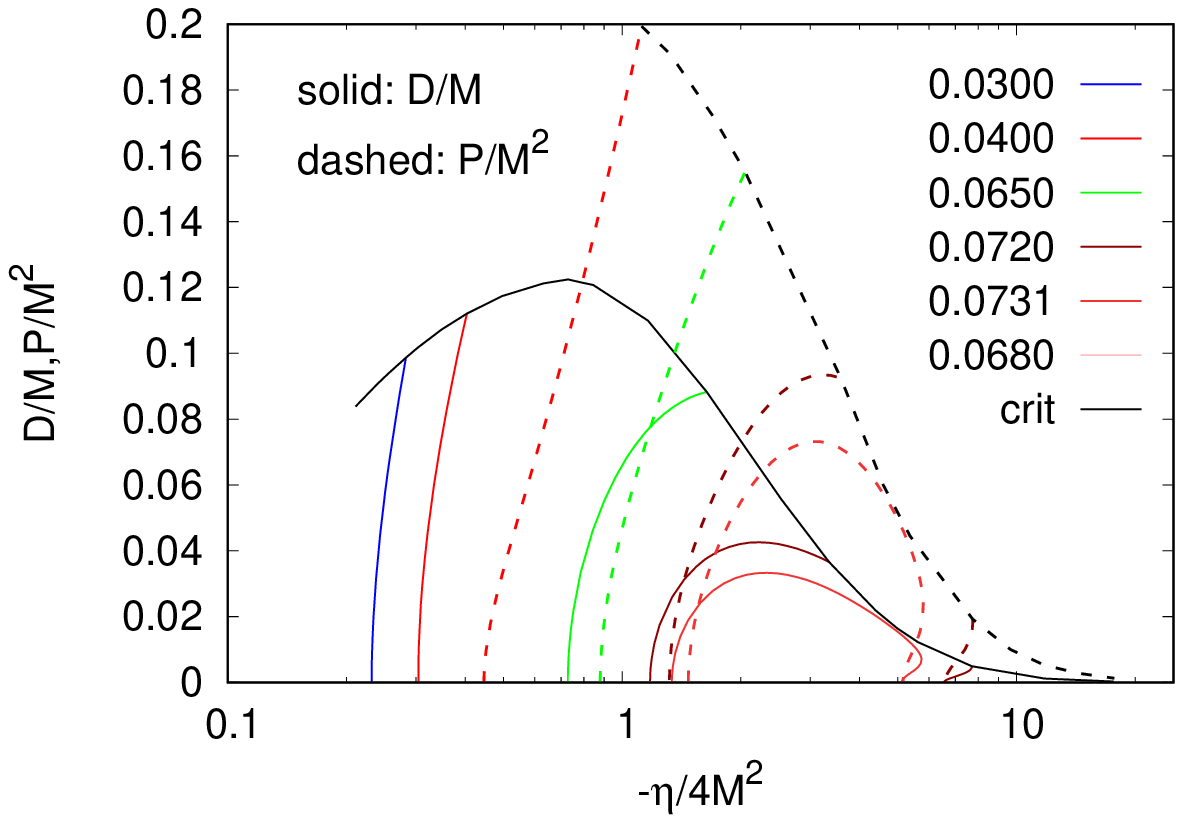}
(b)\includegraphics[width=.45\textwidth, angle =0]{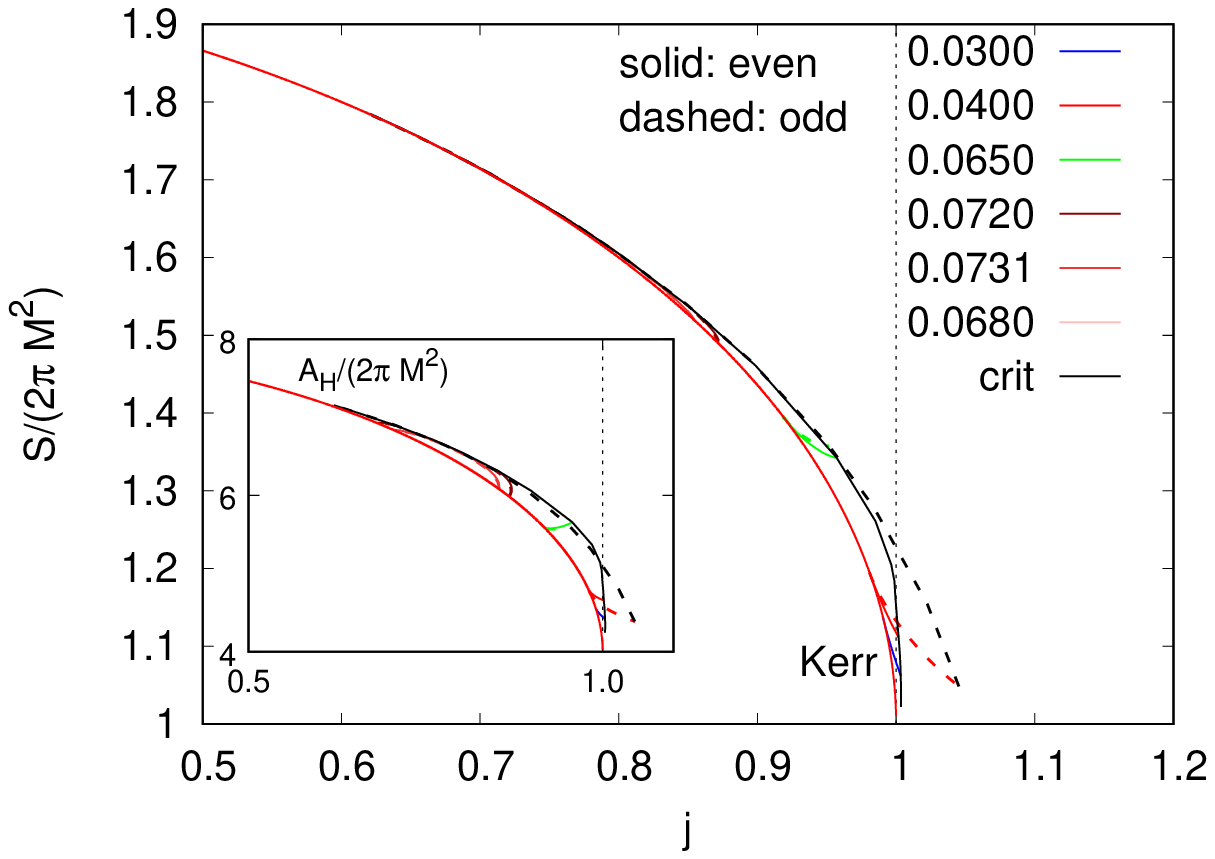}
%(a)\includegraphics[width=.45\textwidth, angle =0]{EsGBBu_dQP_vs_a.eps}
%(b)\includegraphics[width=.45\textwidth, angle =0]{EsGBBu_dSandQ_vs_j.eps}
\end{center}
\caption{Rotating EsGB black holes $(\eta<0)$: (a) scaled scalar charge $D/M$
and scaled dipole charge $P/M$ vs scaled
coupling constant $-\eta/4M^2$,
(b) scaled entropy $S/2\pi M^2$
vs scaled angular momentum $j$, with
scaled horizon area $A_H/16 \pi r_H^2$ vs $j$ in the inset.
}
\label{Fig7}
\end{figure}

Following this interesting observation the associated rotating scalarized black holes were constructed in
\cite{Herdeiro:2020wei} for the exponential coupling function
and in \cite{Berti:2020kgk} for the quadratic one.
Whereas the exponential coupling function did not yield any surprises,
the quadratic one did.
Namely for the spin induced rotating scalarized black holes
the entropy is larger than the Kerr entropy also for the simple quadratic coupling.
Therefore these solutions could be stable, as well.
%The (mathematical) reason is of course, that the additional term
%in the entropy from the Gauss-Bonnet term
%now enters with another sign and thus leads to an increase.
We illustrate the domain of existence of the spin induced
rotating scalarized black holes with quadratic coupling in Fig.~\ref{Fig7}.

Already for the onset of the scalarization several different modes were studied  \cite{Dima:2020yac},
where besides even parity modes also odd parity modes were included
(which do not exist in the spherically symmetic case, of course).
In the two parity sectors the scalar field transforms as
$\varphi(\pi-\theta) = + \varphi(\theta)$ and
$\varphi(\pi-\theta) = - \varphi(\theta)$, respectively.
The fundamental rotating scalarized black holes have even parity
and a monopolar scalar field at infinity,
whereas the odd parity black holes represent excited solutions
whose lowest term at infinity is a dipole term.
Therefore one can associate a monopole or scalar charge $D$  to the even parity solutions,
and a dipole charge $P$ to the odd parity solutions.

In Fig.~\ref{Fig7}(a) we show the scaled scalar charge $D/M$
and the scaled dipole charge $P/M$ versus the scaled
coupling constant $-\eta/4M^2$ for the quadratic coupling function
and no self-interaction  \cite{Berti:2020kgk}.
Both types of solutions represent the lowest solutions
in their respective parity sector.
At vanishing scalar and dipole charge,
the bifurcation from the Kerr solutions takes places.
The critical line then marks the upper boundary of the domain
of existence of even ($D/M$) and odd ($P/M$) 
rotating scalarized black holes. 
The various curves threading the domain of existence
correspond to fixed values of the horizon angular velocity
of the black holes.

Fig.~\ref{Fig7}(b) demonstrates that the parity is indeed 
larger for these rotating scalarized black holes than for the 
Kerr black holes. Here the scaled entropy $S/2\pi M^2$
is shown versus the scaled angular momentum $j$,
again for the lowest solutions in both parity sectors.
Clearly, the Kerr bound $j \le 1$ can be violated for 
such rapidly rotating scalarized black holes.
The inset of the figure shows the 
scaled horizon area $A_H/16 \pi r_H^2$, for comparison,
for both parity sectors.

\section{Conclusions}

Among the numerous alternative theories of gravity
EdGB and EsGB theories are theoretically very attractive,
since they are motivated from quantum gravity theories,
possess second order field equations,
and avoid Ostrogradski instabilites and ghosts.
Here a dilaton or a general scalar field is coupled
to the Gauss-Bonnet term, which is quadratic in curvature.
While there are already significant constraints
on EdGB theories, EsGB theories are much less constrained.

Black holes in EdGB theories have been studied since the
nineties, first the static black holes and later the
rotating black holes. Because of the specific
dilatonic coupling of the scalar field to the Gauss-Bonnet
term, all black hole solutions in EdGB theories carry
dilatonic hair, while the GR black holes do not solve
the set of EdGB field equations.

This is different for EsGB theories, when the coupling
function satisfies appropriate conditions.
Then the GR black holes remain solutions of the
EsGB field equations. However, they undergo 
tachyonic instabilities, where branches of 
curvature induced scalarized black  holes arise.
In the rotating case even two types of scalarized
black holes are present, those with a static limit,
and those, that exist only for rapid rotation,
which are called spin induced EsGB black  holes.

Since these EsGB theories have
so far survived the constraints that have emerged in the GW emission during
binary mergers, when the scalar coupling function allows for a vanishing
scalar field in the cosmological context, and thus leads
to the same cosmological solutions as the standard cosmological $\Lambda$CDM
model \cite{Sakstein:2017xjx},
this makes them attractive also for dynamical numerical relativity studies.
Recently several groups have already done work in this direction, studying, e.g.,
dynamical scalarization and descalarization in binary BH mergers,
dynamics of rotating BH scalarization, or
dynamical formation of scalarized BHs through stellar core collapse
\cite{Witek:2018dmd,Witek:2020uzz,Silva:2020omi,Kuan:2021lol,Doneva:2021dqn,East:2021bqk}.

\section*{Acknowledgement}

We would like to thank our collaborators: Emanuele Berti, Vitor Cardoso, Lucas G. Collodel, Daniela D. Doneva, Valeria Ferrari, Leonardo Gualtieri, Sarah Kahlen, Panagiota Kanti, Fech Scen Khoo, Caio F. B. Macedo, Sindy Mojica, Petya Nedkova, Paolo Pani, Eugen Radu, Kalin V. Staykov, Stoytcho S. Yazadjiev. We gratefully acknowledge support by the
DFG Research Training Group 1620 {\sl Models of Gravity}
and the COST Actions CA15117 and CA16104. JLBS would like to acknowledge support from FCT
project PTDC/FIS-AST/3041/2020.

\end{document}